\documentclass[fleqn,usenatbib]{mnras}

\usepackage{newtxtext,newtxmath}

\usepackage[T1]{fontenc}
\usepackage{ae,aecompl}

\usepackage{graphicx}	
\usepackage{amsmath}	
\usepackage{amssymb}	

\newcommand{\Msun}{\,\ensuremath{\mathrm{M}_\odot}}
\newcommand{\FeoH}{\,[\ensuremath{\mathrm{Fe}/\mathrm{H}}]}
\newcommand{\EuoH}{\,[\ensuremath{\mathrm{Eu}/\mathrm{H}}]}
\newcommand{\EuoFe}{\,[\ensuremath{\mathrm{Eu}/\mathrm{Fe}}]}

\title[R-process elements in the UFDs]{R-process enrichment in ultrafaint dwarf galaxies}

\author[Y. Tarumi et al.]{
Yuta Tarumi,$^{1}$\thanks{E-mail: ytarumi@utap.phys.s.u-tokyo.ac.jp}
Naoki Yoshida,$^{1,2,3}$
and Shigeki Inoue$^{1,2,4,5}$
\\
$^{1}$Department of Physics, School of Science, The University of Tokyo, Bunkyo, Tokyo 113-0033, Japan\\
$^{2}$Kavli Institute for the Physics and Mathematics of the Universe (WPI), UTIAS, The University of Tokyo, Chiba 277-8583, Japan\\
$^{3}$Research Center for the Early Universe, School of Science, The University of Tokyo, Bunkyo, Tokyo 113-0033, Japan\\
$^{4}$Center for Computational Sciences, University of Tsukuba, Ten-nodai, 1-1-1 Tsukuba, Ibaraki 305-8577, Japan\\
$^{5}$Chile Observatory, National Astronomical Observatory of Japan, Mitaka, Tokyo 181-8588, Japan
}

\date{Accepted XXX. Received YYY; in original form ZZZ}

\pubyear{2020}

\begin{document}
\label{firstpage}
\pagerange{\pageref{firstpage}--\pageref{lastpage}}
\maketitle

\begin{abstract}
  We study the enrichment and mixing of r-process elements in ultrafaint dwarf galaxies (UFDs). 
  We assume that r-process elements are produced by neutron-star mergers (NSMs),
  and examine multiple models with different natal kick velocities and explosion energies. 
  To this end, we perform cosmological simulations of galaxy formation to follow mixing of
  the dispersed r-process elements driven by 
  star formation and the associated stellar feedback in progenitors of UFDs.  
We show that the observed europium abundance in Reticulum II is reproduced 
by our inner explosion model where a NSM is triggered at the centre 
of the galaxy, whereas the relatively low abundance in Tucana III 
is reproduced if a NSM occurs near the virial radius of the progenitor galaxy.
The latter case is realised only if the neutron-star binary
has a large natal kick velocity and 
travels over a long distance of a kilo-parsec before merger.
In both the inner and outer explosion cases, it is necessary for the 
progenitor galaxy to sustain
prolonged star formation over a few hundred million years after the NSM, so that
the dispersed r-process elements are well mixed within the inter-stellar medium. 
Short-duration star formation results in inefficient mixing, and
then a large variation is imprinted in the
stellar europium abundances, which is inconsistent with 
the observations of Reticulum II and Tucana III.

\end{abstract}







\section{Introduction}
    
Elements heavier than iron are mainly synthesised by neutron-capture processes 
that occur in neutron-rich environments. The processes are divided 
into `r-process' and `s-process' by the neutron density of the production site.
The s-process is thought to take place in stars in their asymptotic giant 
branch phases, whereas the major astrophysical r-process site is still under debate \citep{Cowan19_Rprocessreview}. 
Neutron-star mergers (NSMs) are the most promising r-process sites,
as have been suggested by recent observations of gravitational waves 
from GW170817 and its electromagnetic counterpart \citep{Abbott_GW170817, Coulter17_EMcounterpart}. The observed electromagnetic spectrum suggests that 
the NSM produces a significant 
amount of r-process elements \citep{Watson2019}.

There have been a number of theoretical studies 
    that consider explosive events such as NSMs and supernovae 
    as astrophysical factories of r-process elements. 
    Numerical relativity simulations show that NSMs can synthesise a 
    large amount of r-process elements (e.g., \citet{2013Hotokezaka_NSM}). 
    Interestingly, however, Galactic chemical evolution models 
    do not favour NSMs as dominant sources 
    \citep{van_de_Voort19_MWRprocess, Safarzadeh19_MWRprocess}. 
    Normal core-collapse supernovae (CCSNe) were thought to be
    another production site, but detailed 
    calculations show that a large flux of neutrinos produced during core collapse
    effectively converts neutrons to protons, rendering r-processes inefficient overall.
    Only a particular type of supernovae may produce a significant 
    amount of r-process elements \citep{Woosley93_collapsar, Siegel19_collapsar, Nishimura15_MRSNe}.
    Since there are likely multiple r-process enrichment channels, 
    it is important to study the chemical signatures of r-process
    elements in galaxies from the early through to the present epoch.

Ultrafaint dwarf (UFD) galaxies are small satellite galaxies 
in the Local Group. The typical total luminosity is less 
than $10^{5} \mathrm{L}_{\odot}$, and the mass is dominated by dark matter. 
UFDs are ideal systems to study chemical evolution as they 
are thought to preserve 
the information on elements produced early in their 
formation histories \citep{Simon19_UFDreview}. 
Recent observations show that two UFDs contain stellar populations with peculiar elemental abundances.
Some stars in Reticulum II (Ret~II) and Tucana III (Tuc~III) show
high europium (Eu) abundances, suggesting
early r-process enrichment events \citep{Ji16_RetII, Hansen17_TucIII},
but it remains unknown why only these two galaxies show clear r-process element
signatures.
    
    \cite{Safarzadeh17_UFD} consider NSMs as a dominant r-process source, 
    and study the effect of different explosion energies
    and delay times of NSMs using cosmological 
    hydrodynamics simulations. 
    They find that neither 
    explosion energy nor delay time significantly affects 
    the overall distribution of Eu within small galaxies, 
    but the environment of the explosion site is important.
    \cite{Safarzadeh19_UFD} 
    further explore the relation between binary population synthesis models 
    and the fraction of r-process enriched UFDs. They compare two binary population synthesis models with different initial binary separation distributions. 
    To reproduce the observed r-process enriched fraction of UFDs of $2/14 \simeq 14$ per cent, the short-separation models are preferred. 
    The merger time is a crucial factor to determine the enriched fraction because the kick velocity is typically higher than the escape velocities of the UFD progenitors. 
    It is found that NSMs contributes to r-process enrichment if the final merger occurs within about 10 per cent of the virial radius of the galaxy. 
     Overall, understanding the differences in the elemental abundances of UFDs 
     help us with identifying the physical conditions of
     r-process enrichment.
     
         In the present paper, we consider NSMs as the major
    source of r-process enrichment in UFDs. We run a set of cosmological 
    simulations of early galaxy formation, and model the r-process enrichment and mixing
    in galaxies that have different star formation histories. 
    We study the dependence of the Eu abundances on star-formation histories 
    by comparing our simulations with recent observations. We examine how neutron-star 
    binary kicks and gas mixing efficiency affect the
    elemental abundance patterns.
    The rest of the paper is organized as follows. In Section 2, we 
    describe our simulations and physical models. 
    In Section 3, we present 
    the main results of our simulations. In Section 4, we 
    discuss the implication for the astrophysical production site of r-process 
    elements. 
    Finally in Section 5, we summarise our results.

\section{Method}

\subsection{Cosmological simulations}
    
     We run cosmological hydrodynamics simulations of galaxy formation
     using the moving mesh code \textsc{arepo} \citep{Springel10_AREPO, Pakmor16_AREPO_improvement, 2019AREPOrelease}. The simulations adopt the Planck 2018
     cosmological parameters \citep{Planck2018}: $\Omega_{m} = 0.315, \Omega_{b} = 0.049, \sigma_{8} = 0.810, n_{s} = 0.965, H_{0} = 67.4 \mathrm{km s^{-1}Mpc^{-1}}$. The basic code settings and physical parameters 
     are the same as those in Auriga simulation \citep{Grand17_Auriga}. 
     Details of the physical models are presented there.
    
    We use the MUlti-Scale Initial Condition generator \textsc{music} \citep{Hahn11_MUSIC} to generate the cosmological initial conditions. 
    The box size is 1 comoving $h^{-1}$Mpc
    on a side. First, we run a low-resolution simulation to locate 
    target haloes (UFD hosts) for zoom-in simulations with higher resolution. In the parent low-resolution simulation, the dark matter particle mass is $6.5\times 10^{3} \Msun$, and the typical gas cell mass is $\sim 1.2\times 10^{3} \Msun$. 
    We select haloes with $\sim 10^{8} \Msun$ at redshift $z = 8$
    as UFD progenitors \citep[see, e.g.,][]{Safarzadeh18_UFDselection}. 
    We then re-simulate three UFD candidates with a higher mass resolution.
    The zoom-in region is an ellipsoidal region enclosing all particles within $2R_{\rm virial}$ of a target halo at the final redshift ($z=6.6$). The mass of each dark-matter (DM) particle is $m_\mathrm{DM} = 102 \Msun$, and the typical gas cell
    mass is $m_\mathrm{g} = 19 \Msun$. The simulations are evolved to $z = 6$ when reionisation proceeds and effectively quenches star formation in UFD progenitors.
    
    We focus on the three haloes (galaxies) that have notably different star formation histories.
    In Fig.~\ref{fig:projection and SFH}, we show the iron abundance  
    and star formation histories of the three galaxies. 
    The average iron abundance of stars is 
    [Fe/H] = -2.35, -2.49, and -2.64 for Halo 1, Halo 2, Halo 3, respectively, 
    which are close to that of Ret~II (-2.65: \cite{2015Simon_RetIIobs}) and Tuc~III (-2.42: \cite{2017Simon_TucIIIobs}). 
    
\subsection{R-process element production by NSM}

    For a UFD progenitor,
    the NSM explosion site affects the overall enrichment level 
    and the spatial distribution of r-process enriched stars, whereas
    the star formation history 
    critically sets the mixing efficiency of the dispersed elements.
    We consider Eu as a major, representative r-process element.  
    We assume that an NSM produces $2\times 10^{-4}\Msun$ of Eu.
    This corresponds to the case where each NSM produces $0.05 \Msun$ of r-process elements \citep{2017Cowperthwaite_NSM} if the abundance pattern is the same as the solar r-process abundance at mass number greater than A$=$90 \citep{Arnould07_solarRprocess}.
    
\begin{figure*}
        \centering
        \includegraphics[width=2\columnwidth]{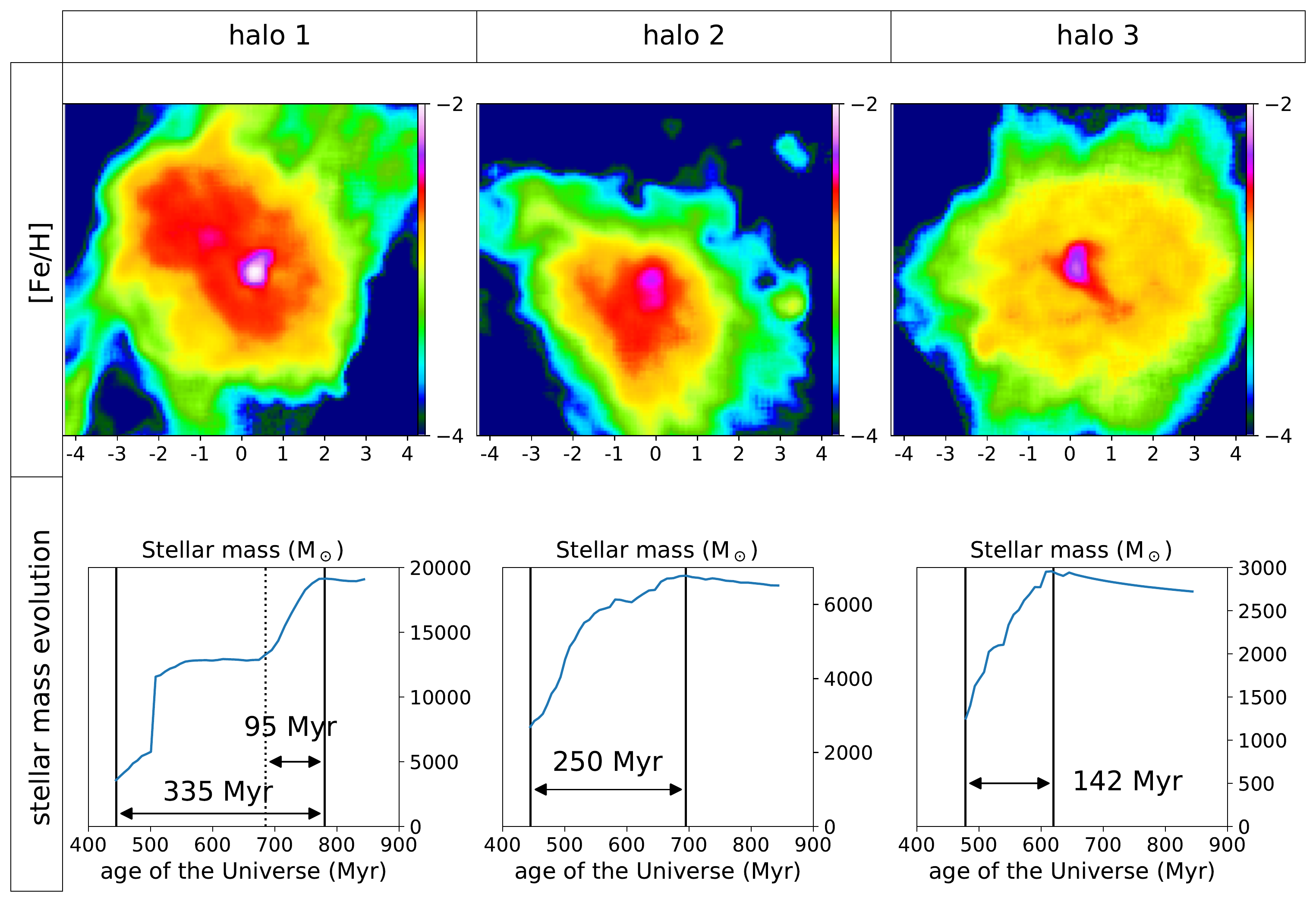}
        \caption{Projected iron abundance and star formation histories. For Halo 1, Halo 2, Halo 3, from left to right. The density-weighted iron abundance is normalised so that the Solar value is 0. The redshift is 6.4 and the numbers on the ticks show the distance from the center in physical kpc.}
        The solid vertical lines in bottom panels show the beginning (dispersal of the r-process elements) and the end of the simulation. The dashed vertical line in the bottom left panel shows the moment of r-process element dispersal in our `late' model.
        \label{fig:projection and SFH}
    \end{figure*}

    It is important to examine the effects of kicks and merger delay times of NS binaries. 
    An NS binary can have a large velocity, comparable to or greater than the escape velocity of the host UFD and the progenitors  \citep[e.g.][]{1997Fryer_DNSkick, Safarzadeh17_NSBkick}.
    It can travel over a long distance, and the final merger can happen outside the star-forming region or even outside the galaxy. 
    Instead of following the orbits of individual NS binaries,
    we model merger events as point explosions at designated locations
    inside and outside the virial radius of the halo. 
    In this way, we can examine how the abundances of r-process elements in metal-poor stars in the UFDs differ in cases with explosions at various locations in a galaxy. 
    In practice, we trigger one NSM at the galactic centre and others at $26\times 8$ points on the concentric spheres with eight different radii $0.1r_{v}, 0.5r_{v}, 1.0r_{v}, 1.5r_{v}, 2.0r_{v}, 2.5r_{v}, 3.0r_{v}, 5.0r_{v}$. With polar coordinates, 26 points can be expressed as: $(\theta, \phi) = (0, 0), (\pi/4, i), (\pi/2, i), (3\pi/4, i), (\pi, 0)$ with $i$ runs every $\pi/4$ from 0 to $7\pi/4$. We flag and trace the ejecta from different explosions independently, and thus are able to examine  different cases in a single simulation run.
    Mixing of the NSM ejecta
    including r-process elements is driven by star-formation and the
    associated feedback effects by supernovae and stellar winds. 
    Thus the relative timing of the onset of star formation and NSMs, 
    and the duration of the subsequent star formation are important.
    We study three galaxies with different star formation histories (Table~\ref{tab:halo_properties} and Fig.~ \ref{fig:projection and SFH}). 
    The explosion time is chosen so that (i) a substantial amount of stars have already been formed, and (ii) the galaxy still produces stars after the NSM. The first condition is necessary to produce the binary neutron-stars, and the second condition is imposed to see the mixing of the r-process ejecta. For the galaxies we study here, the total gas mass decreases towards the final output redshift, $z=6.6$. This is owing to photo-heating and evaporation caused by hydrogen reionisation, and also to outflows driven by stellar feedback. Low-mass haloes like UFD progenitors have shallow gravitational potential wells and thus cannot retain the photo-heated gas after reionisation. Star formation is quenched at this point.
    \begin{table}
        \centering
        \begin{tabular}{|l|c|c|}
         & stellar mass after NSM(\Msun) & SF duration(Myr) \\
        \hline
        Halo 1 & 15000 & 335\\
        Halo 2 & 4000 & 250\\
        Halo 3 & 1500 & 142\\
        Halo 1, late & 5000 & 95\\
        \hline
        Ret~II & 2600 & - \\
        \hline
        \end{tabular}
        \caption{Halo properties in terms of star formation. The stellar mass of Ret~II is the observationally inferred value at $z=0$. \citep{Bechtol15_RetIIobservation}}
        \label{tab:halo_properties}
    \end{table}

    We model the NSM bubble evolution by following analytically a point explosion in the self-similar through to snow-plough phases \citep{GalaxyFormaitonAndEvolution}. We calculate the radius and velocity in the self-similar phase as a function of $t$ assuming the explosion energy of $E = 10^{51}$ erg.
    The end point of the self-similar phase is assumed to the time when a quarter of the explosion energy is radiated away. By then, the blast-wave reaches 
    \begin{equation}
        r_\mathrm{sh} = 23\times \biggl(\frac{n}{1\mathrm{cm^{-3}}}\biggr)^{-19/45}\times \biggl(\frac{E}{10^{51}\mathrm{erg}}\biggr)^{13/45} \mathrm{pc}
    \end{equation}
    with the velocity of
    \begin{equation}
        v_\mathrm{sh} = 200\times \biggl(\frac{n}{1\mathrm{cm^{-3}}}\biggr)^{2/15}\times \biggl(\frac{E}{10^{51}\mathrm{erg}}\biggr)^{1/45} \mathrm{km~s^{-1}},
    \end{equation}
    where $n$ is the number density of hydrogen atoms in unit of $cm^{-3}$. 
    Afterwards, the shocked shell expands while conserving momentum
    with the velocity scaling as $v \propto r^{-3}$.
    We assume that the snowplough phase ends when the shell velocity decreases to the value comparable to
    the turbulent velocity of the surrounding gas ($\sim 10 \mathrm{km~s^{-1}})$. 
    The final radius $r_{\rm sp}$ is calculated as
    \begin{equation}
        r_\mathrm{sp} = r_{\rm sh}\times \biggl(\frac{v_{\rm sh}}{10\mathrm{km~s^{-1}}}\biggr)^{1/3}.
    \end{equation}
    In each simulation, we define the `NSM bubble' with the radius $r_{\rm sp}$. 
    For the gas cells inside $r_{\rm sp}$, we distribute r-process elements with weighting by the cell volume. 
    Effectively, we assume that the ejecta is well mixed within the shell, but
    also approximately model that the dense (small volume) cells are less 
    susceptible to enrichment. 
    
    We have also run simulations with different explosion energies of $10^{50}$ and $10^{52}$ erg, to obtain qualitatively similar results.
    We find that the main physical process of ejecta dilution is large-scale turbulence
    driven by star formation and galaxy assembly\footnote{We note that the rapid increase at $t_{\rm age} \sim 500$ Myr of the stellar mass of
    Halo 1 shown in Figure 1 is caused by mergers of progenitor galaxies.}, 
    and the exact bubble size at the snowplow phase does
    not significantly affects the dilution efficiency.

\section{Results}

We compare the stellar Eu abundances in our simulated galaxies
with those of the two UFDs, Tuc~III and Ret~II.
We then identify models that reproduce the observed features.

\begin{figure}
	\includegraphics[width=\columnwidth]{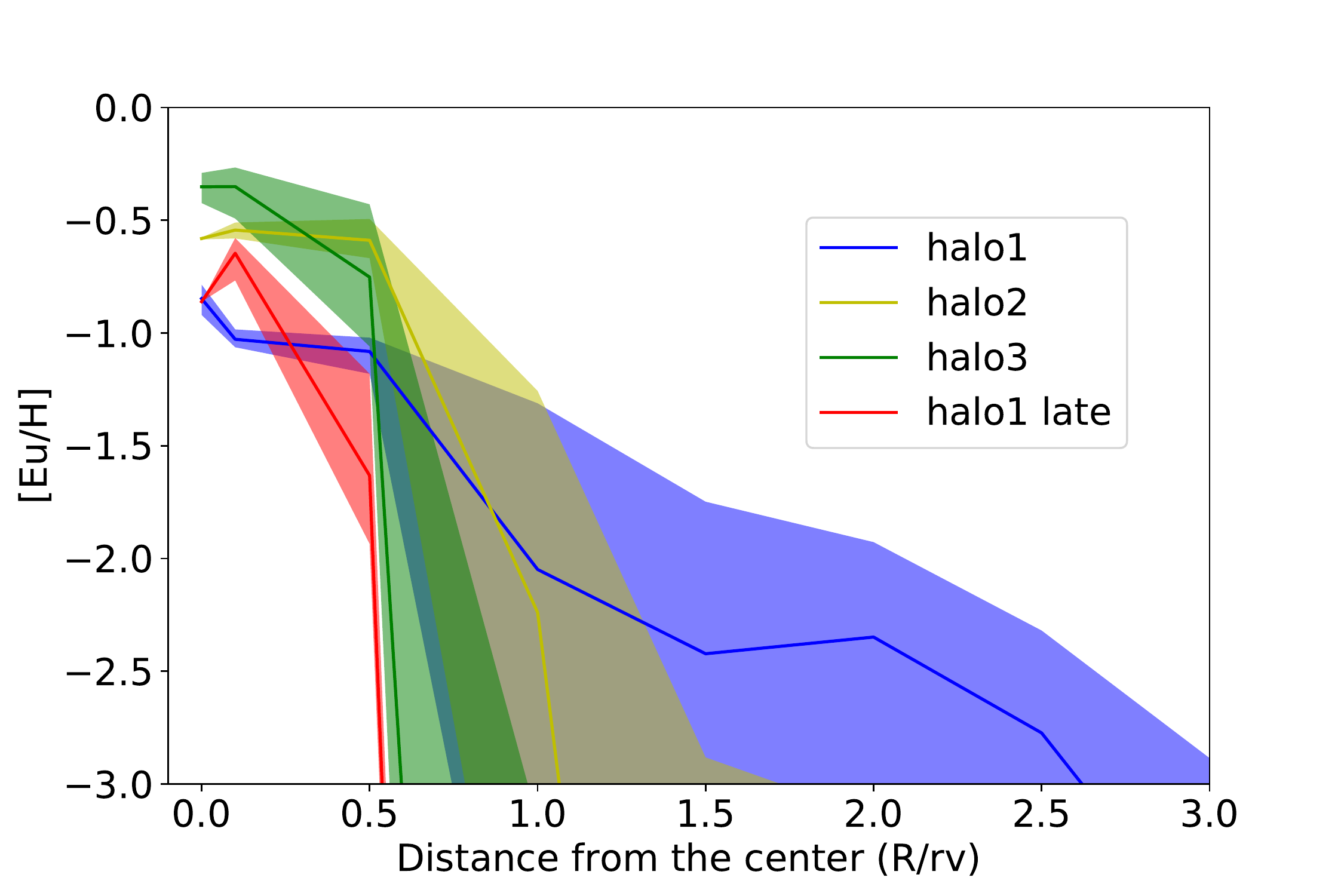}
	\caption{The mean [Eu/H] of stars as a function of distance of the explosion site from the galaxy centre. The coloured regions are representing 25 percentile and 75 percentile to show the scatter in each radius.
	}
    \label{fig:Overall abundance}
\end{figure}

\subsection{NS merger explosion site}
\begin{figure*}
    \centering
    \includegraphics[width=\columnwidth]{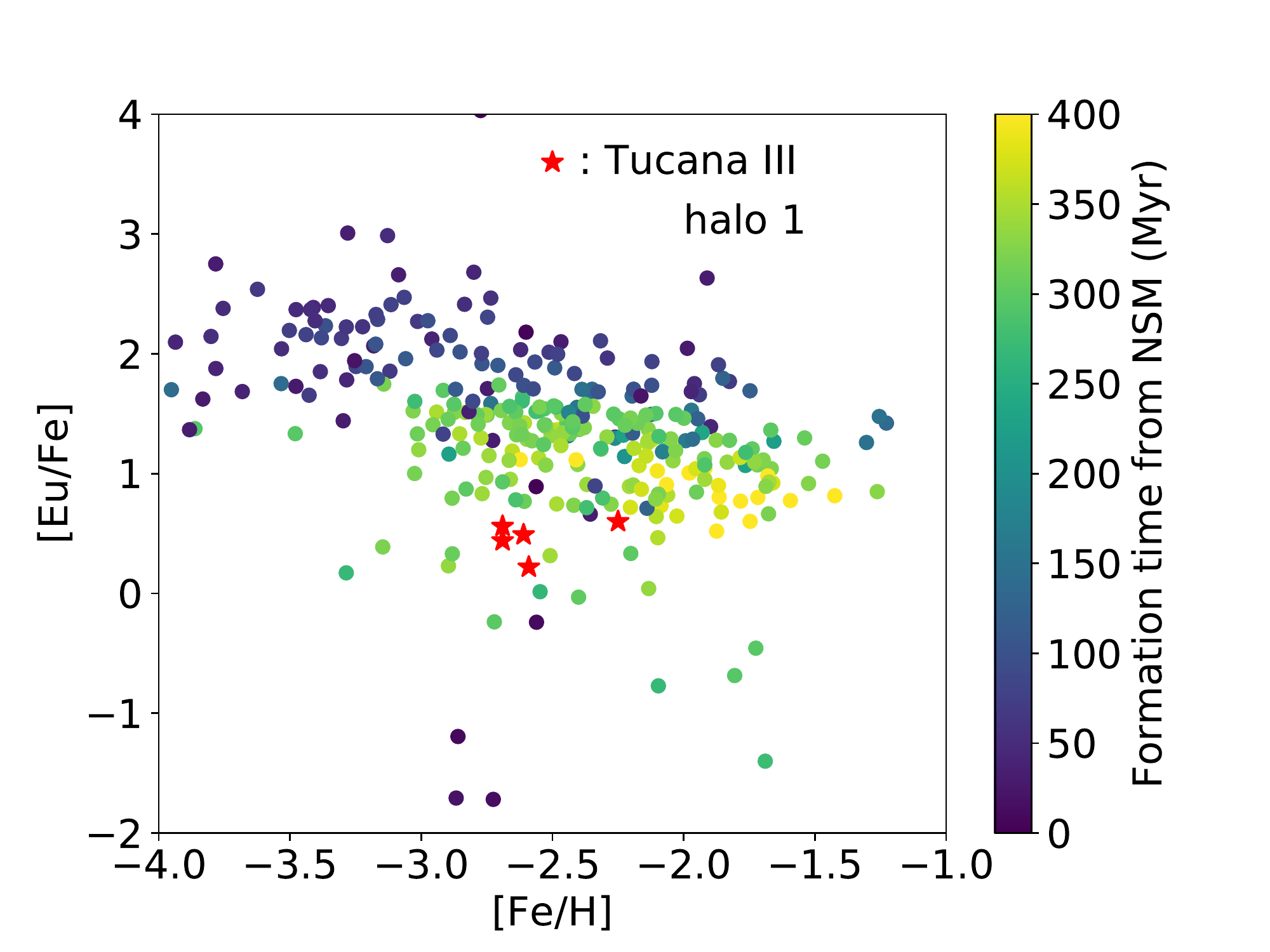}
    \includegraphics[width=\columnwidth]{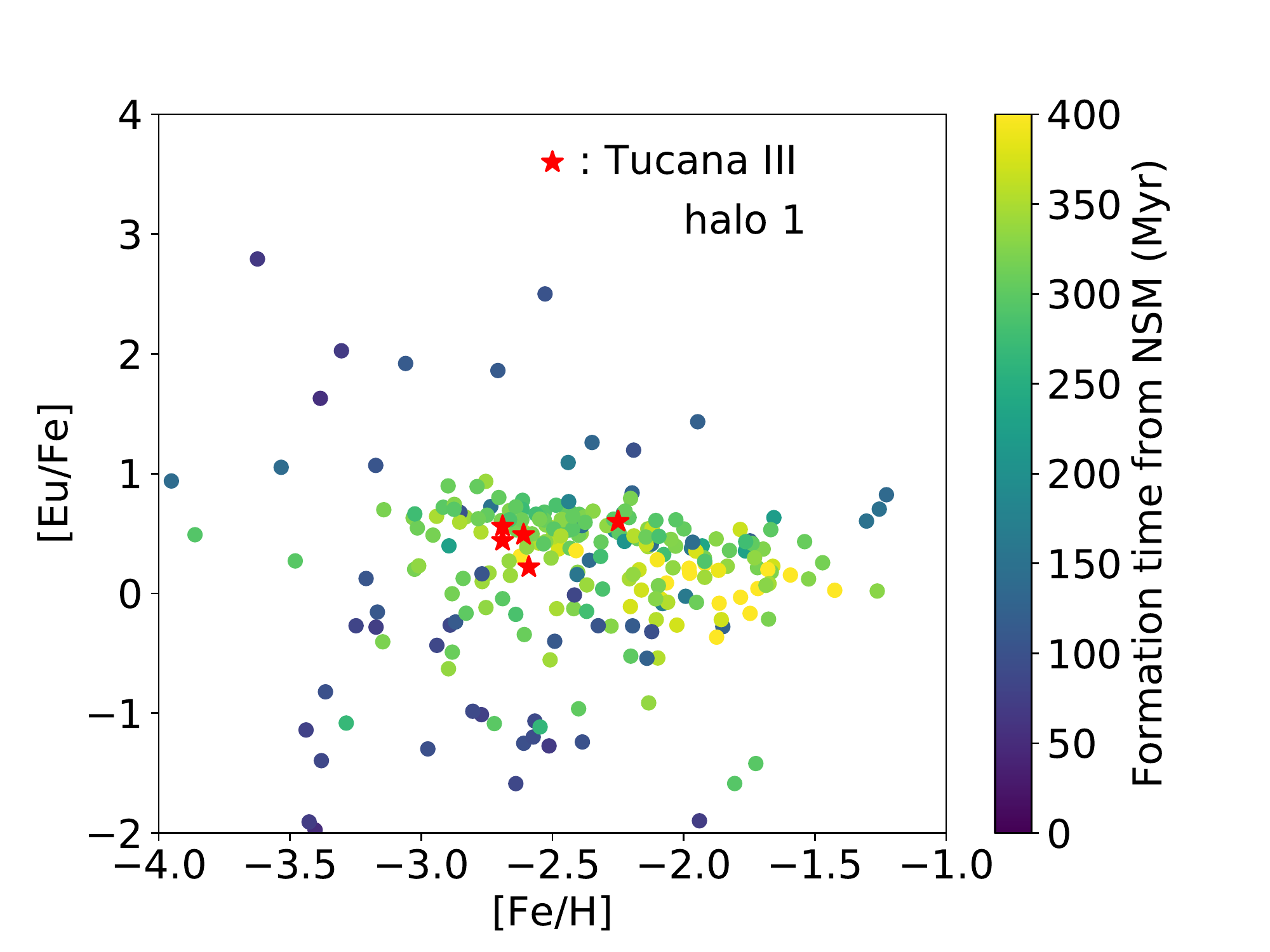}
    \caption{Stellar [Eu/Fe] - [Fe/H] abundances for halo 1, 
    compared with the observational data of Tuc~III.
    Left panel: The NSM is triggered at the galactic centre where the main star-forming region is located.
    Right panel: The NSM occurs at around virial radius of the galaxy. The mean [Eu/Fe] abundance in the right panel is closer to Tuc~III.}
    \label{fig:Tuc III scatter plot}
\end{figure*}
\begin{figure*}
    \centering
    \includegraphics[width=\columnwidth]{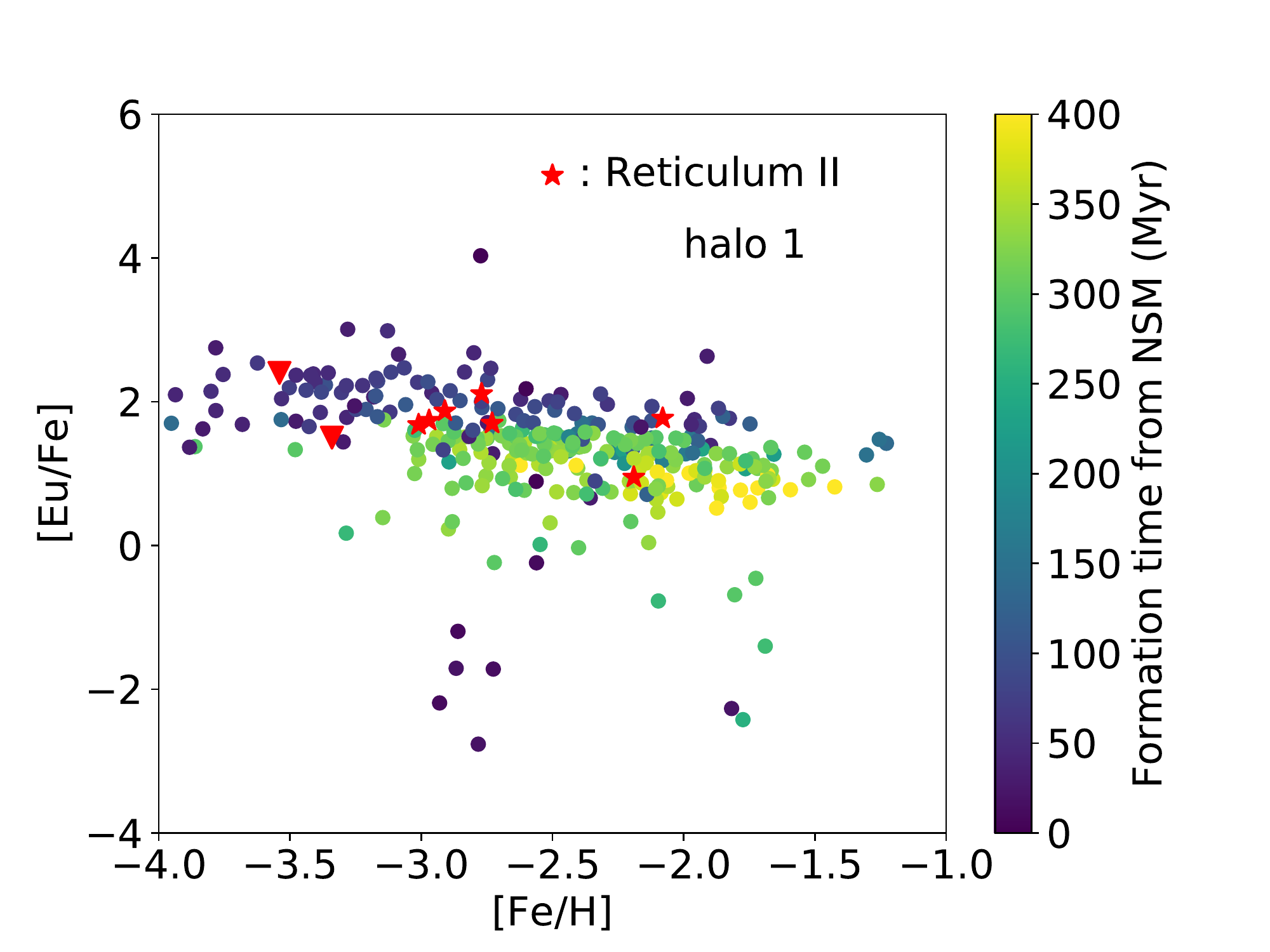}
    \includegraphics[width=\columnwidth]{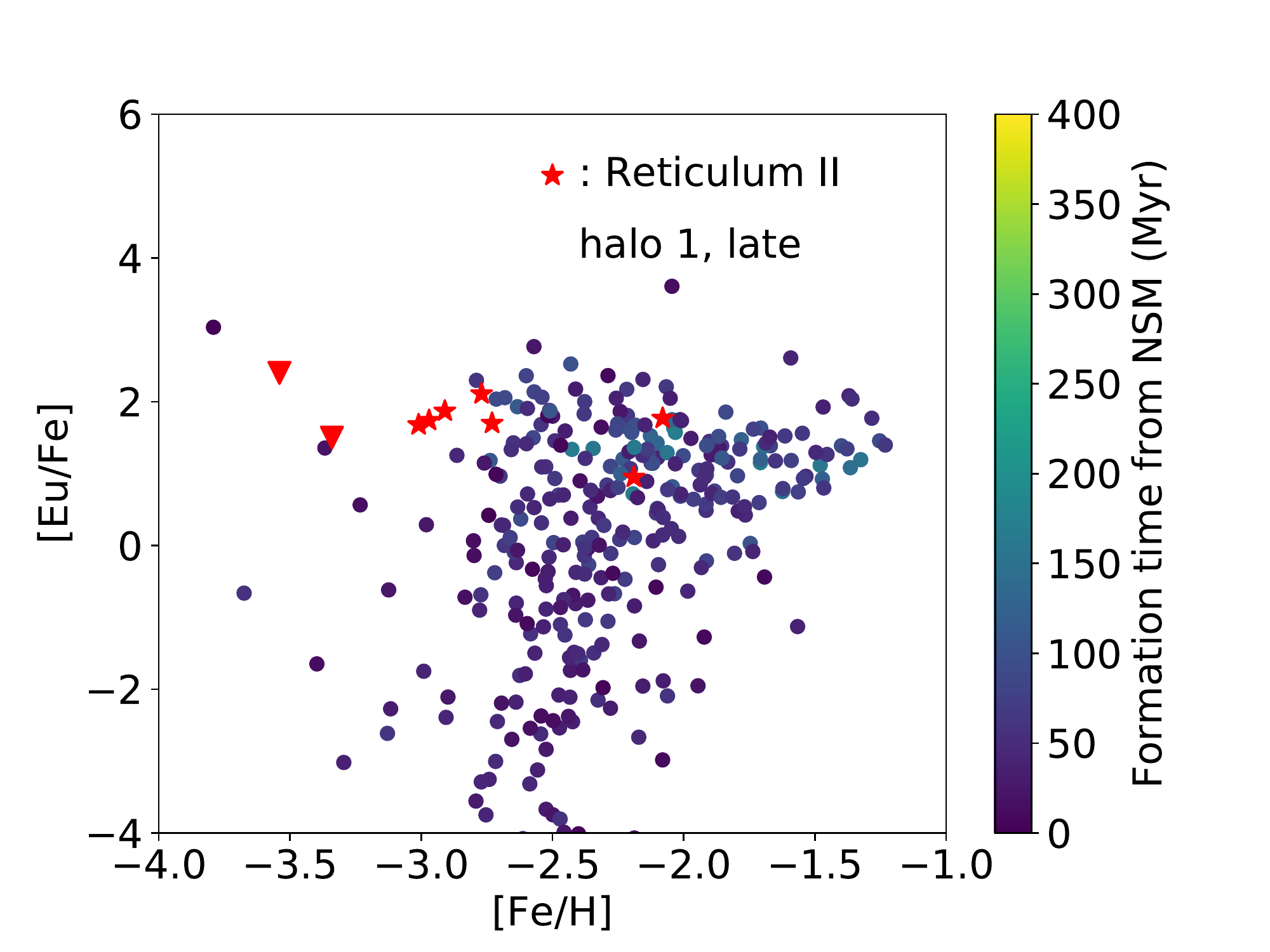}
    \caption{Stellar [Eu/Fe] - [Fe/H] abundances.  
    The NSM is triggered at the galactic centre in these two cases.
    Left: There is a period of long (335 Myr) star formation after the NSM. 
    Right: Star formation is quenched soon (95 Myr) after the NSM explosion. 
    The right panel shows a significantly large scatter of [Eu/Fe] among the member stars, which is not found in the observations of Ret~II.}
    \label{fig:Ret II scatter plot}
\end{figure*}

Fig.~\ref{fig:Overall abundance} shows the mean Eu abundances as a function of the location (distance) of explosion site from the centre. 
As is expected, the mean Eu abundances are progressively lower 
in the cases where a NSM occurs in the outer region at $R > r_{\rm vir}$. 
Interestingly, we see relatively small differences in [Eu/Fe]
if a merger occurs inside the virial radius;
all the three haloes show $\EuoH = -0.5 \sim -1.0$.
Note that the [Eu/H] is as large as the mean abundance of Ret~II (-0.82 if we exclude the stars with non-detection, and -0.93 if we include them). 
Since we consider stochastic r-process enrichment, 
essentially by a single event, \EuoH\ is not
correlated with the stellar mass 
of the galaxy, but is more correlated with the available gas mass
to be mixed.

Tuc~III has a low value of $\EuoH \sim -2.0$ \citep{Marshall18_TucIIIobservation}, which may
suggest either 
(i) NSM explosion(s) took place outside the virial radius, or 
(ii) the gas mass was very large. 
In the former case, the distance from the galactic centre 
should not be too large (Fig.~\ref{fig:Overall abundance}), whereas in the latter,
mixing with a large amount of gas naturally reduces \EuoH. 
To reconcile the value of Tuc~III,
we need 10 times more gas.
Since the baryon mass fraction in the simulated galaxies are consistent to cosmic mean within a factor of two (for other references, see e.g. Fig.4 of \cite{2012Wise}),
the progenitor halo mass should also be larger by a factor of ten, 
to be $\sim 10^{9} \Msun$.
This is too large for star formation in the UFD progenitor 
to be quenched by cosmic reionisation \citep{Safarzadeh18_UFDselection}. 
We thus argue that the outer explosion (above case [i]) is 
favoured over the large mixing mass.

We note that the small scatter in [Eu/Fe] in Tuc~III is not
reproduced in Halo 2 and Halo 3 although the mean [Eu/Fe] 
can be close to that of Tuc~III.
Halo 2 and Halo 3 show significantly large scatters as a 
result of inefficient mixing of r-process elements 
due to the short star formation duration. 
We discuss this point further in the next section.

\subsection{Star formation history}
    
    An interesting feature of the two UFDs is that the stellar [Eu/Fe] is
    roughly constant, with little or only weak dependence on [Fe/H] for Ret~II.
    Also the overall scatter in \EuoFe\ is small \citep{Ji16_RetII},
    and the same is true for Tuc~III \citep{Marshall18_TucIIIobservation}.  
    A successful enrichment model should therefore reproduce or explain 
    such a trend. 
    
    Fig.~\ref{fig:Ret II scatter plot} compares the r-process element abundances 
    of stars in Halo 1 with different explosion timing. Halo 1 experiences a prolonged 
    star formation over 335 Myr, and the r-process elements dispersed by
    a NSM at en early epoch are well mixed within the inter-stellar medium. 
    Consequently,
    the scatter in [Eu/Fe] becomes fairly small. 
    The [Eu/Fe] - [Fe/H] distribution is remarkably similar to Ret~II. 
    Contrastingly, if the star formation lasts only 95 Myr after the NSM,
    the stellar [Eu/Fe] shows a large scatter. Clearly, efficient 
    mixing of the gas and the dispersed
    heavy elements is preferred or even necessary in order to reproduce the 
    observed small scatter of \EuoFe.
    This implies that Ret~II once had a prolonged star formation activity
    over a few hundred million years before the star formation ceased completely.
        
    Fig.~\ref{fig:Ret II scatter plot} also shows the formation time of stars measured from the moment of NSM explosion (colour-coded as in the indicator on the right of each panel). Both in Figs.~\ref{fig:Tuc III scatter plot} and \ref{fig:Ret II scatter plot}, we 
    see "evolution" in [Fe/H] as time elapses, with slight downwards tilts. 
    The trend reflects the fact that the r-process production event happens once, and [Eu/H] varies little, but the metallicity (iron abundance) increases 
    as the galaxy chemically 
    evolves over time.

    We have shown that mixing of r-process elements critically affects the stellar
    elemental abundances. 
     In our simulations, supernova-driven galactic winds stir the ISM in the galaxy,
     providing a major mixing mechanism.
     Technically, the stellar feedback effect within a star-forming gas cell is modeled by ejecting a wind particle in a random direction \citep{Springel03_WindParticles}. The wind particle carries 40 per cent of the heavy elements that have been synthesised and expelled by the SNe, and the other 60 per cent is immediately distributed to nearby gas cells \citep{Vogelsberger13_MetalStripping}. The wind particle travels until the designated maximum travel time, or until reaching a low-density gas cell with $\rho_\mathrm{g} < 0.05\rho_\mathrm{th}$. The wind particle finally deposits its mass, metal content, momentum, and energy into the gas cell in which it is located at the final moment.

    In the centres of the galaxies, galactic winds efficiently stir the ISM, 
    whereas in the outskirt, full-scale mixing can be achieved only through the halo growth
    via mergers and accretion. 
    We evaluate the typical mixing timescale is 100 Myr
    for all three halos, which is a few times shorter than in the outskirt explosion 
    cases presented in the previous section. 
    The sufficient mixing of elements makes the [Eu/Fe]-[Fe/H] distribution in the central explosion model to be consistent with Ret~II.
    
    \begin{figure}
        \centering
        \includegraphics[width=\columnwidth]{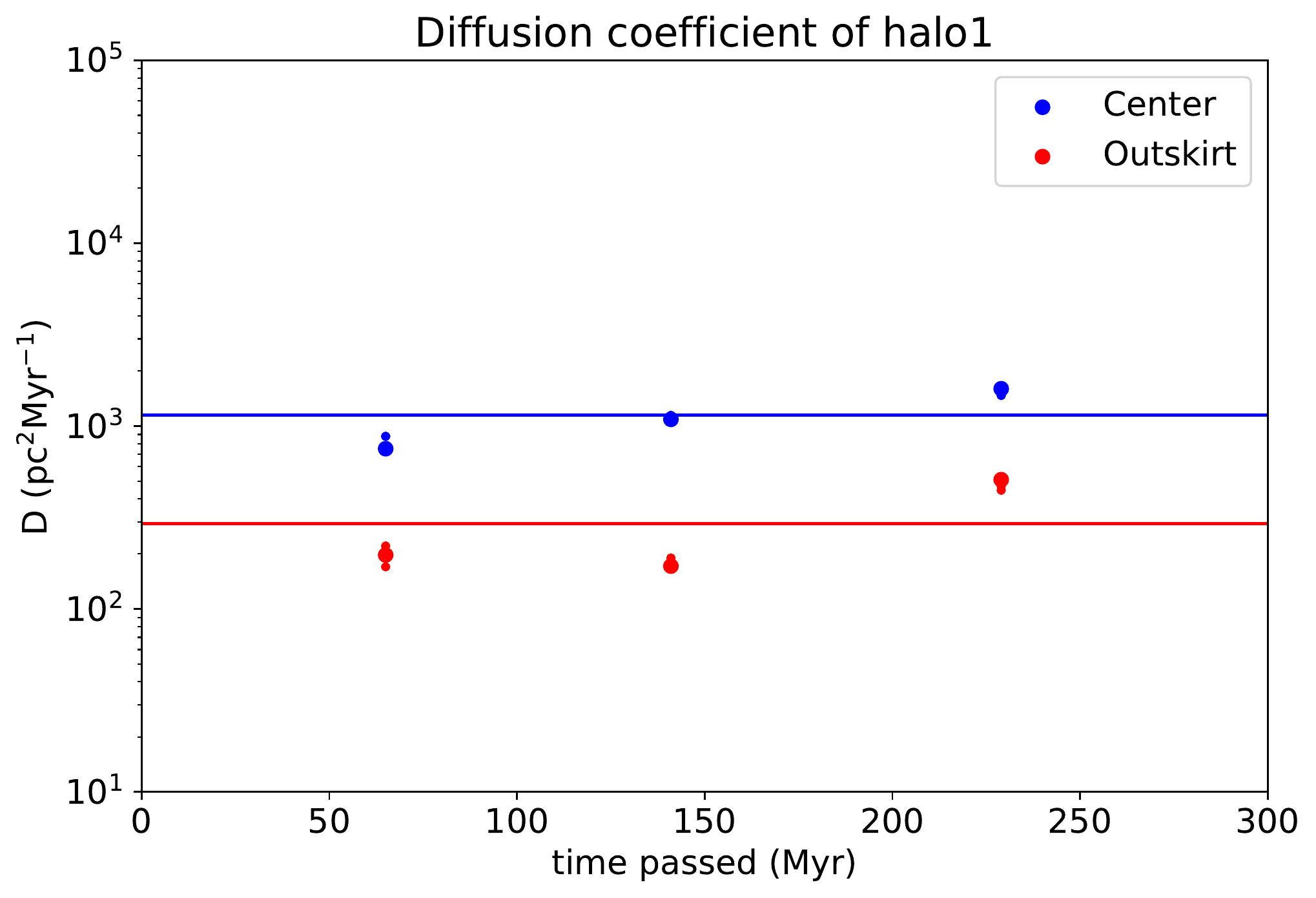}
        \caption{The diffusion coefficient of the r-process elements
         estimated by three-dimensional Gaussian fit
         to the r-process element distribution. The details are described in the
         main text.
         The solid lines indicate the average of the coefficients measured 
         at different times. Large dots show the result from our main simulation. Two small dots at each moment show the results of the same analysis on simulations with eight times better and eight times worse mass resolutions.
         }
        \label{fig:diffusion coefficient}
    \end{figure}
    
    In Fig.~\ref{fig:diffusion coefficient}, we show the diffusion coefficients of Halo 1, obtained by three-dimensional Gaussian fittings of the r-process element distribution. The diffusion coefficient $D$ can be calculated by inverting $2 D t = \sigma^{2}$, where we derive Gaussian standard deviation $\sigma$ by the fitting. We estimate the diffusion coefficients inside the UFD progenitors to be $1 \times 10^{-3}~ \mathrm{kpc^{2}~Myr^{-1}} \simeq 3\times 10^{26} ~\mathrm{cm^{2}~s^{-1}}$, and the one at the outskirt of the galaxy to be $3\times 10^{-4} ~\mathrm{kpc^{2}~Myr^{-1}} \simeq 9\times 10^{25} ~\mathrm{cm^{2}~s^{-1}}$. As a convergence test, we have performed the same analysis to two additional simulations with eight times lower and higher mass resolutions. We have confirmed that the star formation histories are nearly identical and also that the diffusion coefficients are consistent with the main results shown in Fig. 5 within 20 percent. Within 250 Myrs, the r-process elements distributed around the centre are well mixed to a half of the virial radius, which encloses star-forming regions. The value is consistent with but a little smaller than that derived by \cite{2005Karlsson_metalmixing} who suggests the diffusion coefficient of $7\times 10^{-4} ~\mathrm{kpc^{2}~Myr^{-1}}$ using a stochastic metal enrichment model, calibrated by the observed metallicity distribution function of low-mass stars in the MW. \cite{2017Hirai_mixingefficiency} study the `mixing efficiency' by comparing the stellar barium abundances of Milky-Way (MW) and dwarf galaxies to their simulated galaxies using a turbulent metal mixing model. They find that $D > 2\times 10^{-5} ~\mathrm{kpc^{2}~Myr^{-1}}$ is required to make the barium abundance consistent. \cite{2015Ji_diffusioncoefficient} use $D = 2.4\times 10^{-3} ~\mathrm{kpc^{2}~Myr^{-1}}$ as their canonical value for the effective diffusion coefficient of the galaxy with a similar mass and a redshift (note the difference in the definition: their effective $D$ corresponds to our $D/3$). All these values are consistent with our result from the high-resolution hydrodynamics simulations. 
    
    \cite{2019Emerick_mixing} follow metal mixing in a dwarf galaxy using Eulerian hydrodynamics simulations. They report that the mixing proceeds slower if rare elements are deposited in the outskirt of the galaxy than in the case
    with deposition near the centre. This trend is also observed in our simulations. They also show that the mixing time can be estimated by the transport timescale \citep{2013Pan_mixing}: 
    \[\tau_\mathrm{trans} = L_\mathrm{G}^{2}/(L_\mathrm{turb} \,v_\mathrm{rms}).\]
    In our case, we estimate the typical size of the dwarf galaxies as $L_\mathrm{G} \sim 0.5~\mathrm{kpc}$, and the characteristic turbulent length and velocity $L_\mathrm{turb} \sim 100 ~\mathrm{pc}, v_\mathrm{rms} \sim 10 ~\mathrm{km/s}$,
    respectively. The former is roughly the size of a NSM bubble when it is in the snow-plough phase (Eq. [3]) and the latter is the turbulent velocity of the amibient gas.
    With these values, we obtain the mixing timescale of about 250 Myr, which is consistent with our simulations where it takes a few hundred Myr to chemically homogenise the galaxy.
    
\section{Discussion}

\subsection{Star formation histories of the UFDs}

    Our results suggest an interesting possibility that the scatter of r-process elements among stars in UFDs can be used as a measure of mixing efficiency that 
    is determined by
    the timing of the r-process enrichment and the star-formation duration in the galaxy. 
    If we assume a continuous star formation model, it takes
    300 Myrs or more after a NSM explosion for the ejecta including r-process elements
    to be mixed well with the ambient ISM.
    This is longer than the free-fall time $t_\mathrm{ff} = \sqrt{\frac{3\upi}{32G\rho} } \simeq 70\mathrm{Myr}$. 
    Such `sufficient' mixing can be achieved if binary neutron-stars are formed in the very early epoch, and the binary merges quickly, {\it and} star formation lasts long. 
    
    These conditions make the single star-burst event scenario rather unlikely, and make us consider prolonged or multiple star-formation epochs.
    We need more UFD observations to make this argument statistically robust. 
    It would be quite interesting to search for the first-generation stars or "earlier generation" stars 
    that contain no r-process elements 
    in Ret~II and Tuc~III,
     because it would then indicate the timing of NSM relative to the onset of major star formation in the galaxies.

\subsection{Natal kick of neutron-star binaries}

\begin{figure}
    \centering
    \includegraphics[width=\columnwidth]{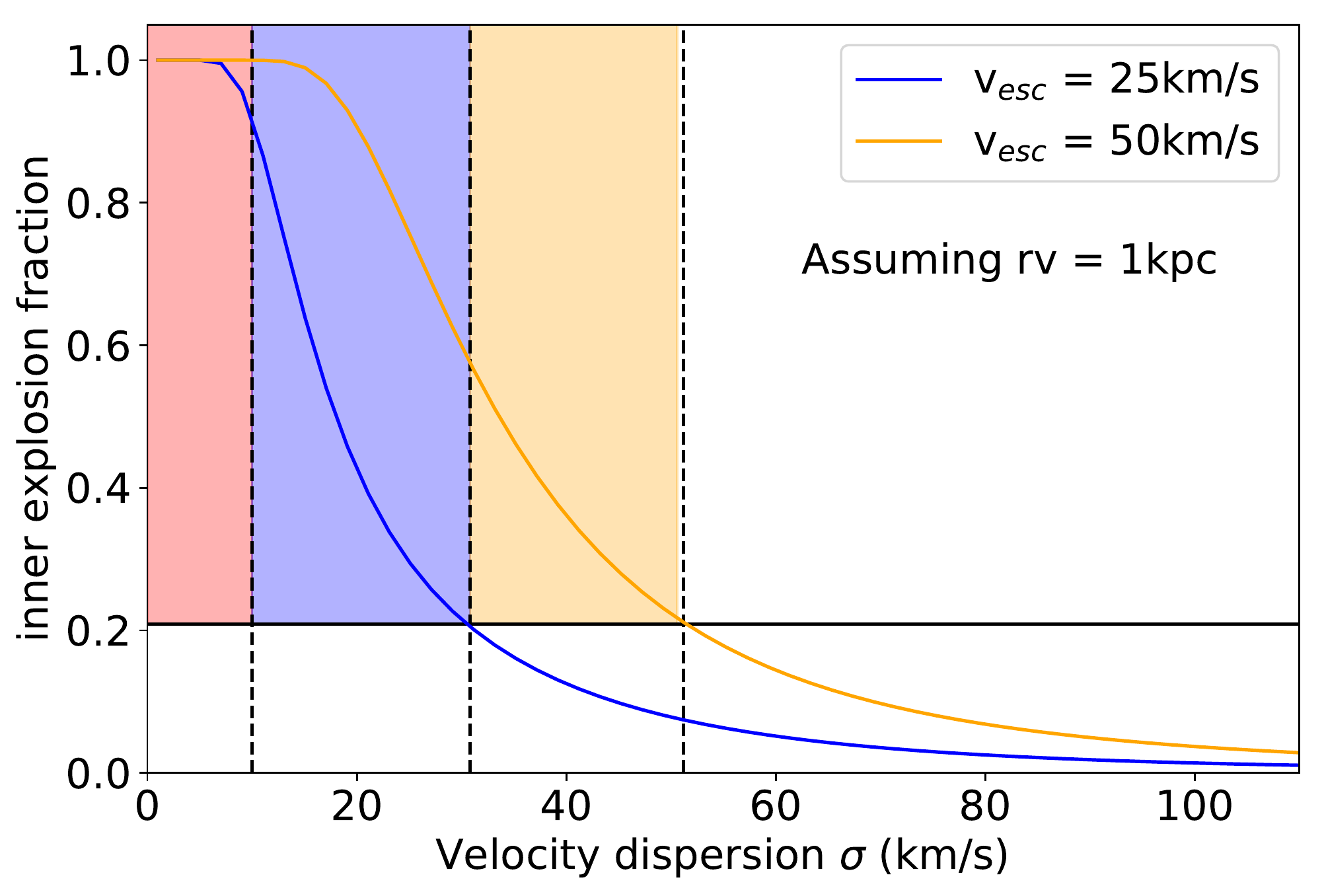}
    \caption{Inner explosion fractions in two haloes with different escape velocities. The horizontal axis indicates the velocity dispersion we have assumed. The horizontal solid line corresponds to the level below which the possibilities of those inner explosion fractions can be excluded with 5 per cent significance. The inner explosion fraction below this line can be excluded with 5 per cent significance. The very low dispersion (less than 10km/s) is unlikely because the instant mass-loss via neutrino kicks the binary neutron-star system at about 10km/s \citep{2016Beniamini_natalkicks}.}
    \label{fig:DTD_kick}
\end{figure}

We have shown that the high Eu abundance in Ret~II 
is reproduced if an NSM occurs at the centre.
Such central explosion can occur if the binary neutron-stars 
receive very weak kick at the formation, or merge within a very short time.
Contrastingly, the relatively low Eu abundance in Tuc~III 
can be explained by NSM explosion in the outer halo beyond the virial radius,
i.e., if the binary neutron-stars receive modest kick comparable to escape velocity of the galaxy, $\sim 25$ km/s depending on the place the binary was born. 
Intriguingly, it is known that 
there are two populations of binary neutron-stars; one is with low peculiar velocities
and the other with high velocities \citep{2016Beniamini_natalkicks}.
It is important to estimate how far a kicked neutron-star binary travels over
a hundred million years.
We calculate the fraction of neutron-star binaries that merge inside the virial radius, assumed to be 1 kpc. 
The initial velocity is drawn from the Maxwell-Boltzmann distribution with a certain velocity dispersion $\sigma$. 
We also assume that the delay time distribution is proportional to $t^{-1}$ \citep{Belczynski18_DTD} with $t_\mathrm{min} = 10~\mathrm{Myr}$ and $t_\mathrm{max} = 10~\mathrm{Gyr}$. 
Fig.~\ref{fig:DTD_kick} shows the resulting fraction as a function of $\sigma$. 
For the assumed merger delay time distribution, the probability is generally 
low that NSMs occur within the virial radius of a UFD, unless the natal kick
velocity is low.
The fraction of UFDs that are enriched with r-process elements
can be used to constrain the typical kick velocity.
Note, however, that at least a few other factors
need to be considered, such as the formation rate of neutron-star binaries and the stellar mass estimate of UFD progenitors. 
Our discussion here is based on simple assumptions as follows:
(i) We assume that the initial mass function IMF in UFDs is Chabrier IMF \citep{Chabrier01_IMF}. 
Since the stellar populations in UFDs are quite old, stars heavier than $0.8 \Msun$ already disappeared or in the post-main-sequence phase.
This reduces the surviving stellar mass to 36 per cent of the initial stellar mass. 
(ii) No tidal stripping takes place, so the initial stellar mass can be derived only from the dead stars correction in (i). 
(iii) The mass-to-light ratio is the same for all UFDs as that of Ret~II \citep{Bechtol15_RetIIobservation}.
There are 14 UFDs whose stellar Eu abundances are measured \citep{Simon19_UFDreview}. With these assumptions, we can estimate the initial stellar mass to be $2.3\times 10^{5} \Msun$. If we adopt the typical fraction of merging neutron-star binaries to $10^{-5}$ per $1 \Msun$ in stars, we expect 2.3 UFDs experience r-process enrichment. 
This is consistent with the number of UFDs enriched ($= 2$), and thus implies that the NSMs should dominantly takes place within the galaxy they were born in. 
The high initial velocity model is disfavoured.

High-velocity NS binaries may not have been able to escape if the UFD progenitor 
was more massive in the past. Interestingly, there are some signatures indicating that the Tuc~III had experienced tidal stripping \citep{Li18_TucIIItidalstrip}. In order to make the model with $\sigma_{\rm kick} > 100~\mathrm{km~s^{-1}}$ viable, the UFD progenitors must lose its original mass more than 90 per cent on average. 

R-process enrichment can be caused by an "external" neutron-star binary
formed in another galaxy. Natal neutron-star kick makes it possible for a binary
to travel over a long distance in a few to several hundred million years.
We find, however, that such an "external enrichment" of UFDs is an unlikely event.
In our parent cosmological simulation, the typical physical distance between two star-forming galaxies is $\sim 60$ kpc at $z \sim 11$, 
and the virial radius of the UFD progenitor is $\sim$ 1 kpc. Then the solid angle subtended by the neighbouring galaxy 
is $\pi/(60)^{2}$, yielding the direction-wise 
success rate of $\sim 10^{-4}$. Furthermore, the NSM must occur
at the right time during its passage through a (small) galaxy; the merger occurring inside a galaxy is roughly (2kpc)/(60kpc) = 1/30. The actual success rate should be even lower because a large fraction of binary neutron-stars explode before traveling over $\sim 60$ kpc. 
In conclusion, we can expect that the external NSM enrichment is extremely rare, and although not entirely
impossible, we can ignore such an enrichment channel.

\begin{table}
    \centering
    \begin{tabular}{|l|c|c|}
         & Time lag < 100Myr & Time lag > 100 Myr \\
        \hline
       R<$r_{v}$ & 
       \begin{tabular}{c}
            High abundance, \\
            large scatter
       \end{tabular} & 
       \begin{tabular}{c}
            High abundance, \\
            small scatter
       \end{tabular}\\
       \hline
       R>$r_{v}$ & 
       \begin{tabular}{c}
            Almost no \\
            r-process elements
       \end{tabular} &
       \begin{tabular}{c}
            Medium abundance, \\
            medium-small scatter
       \end{tabular}\\
       \hline
    \end{tabular}
    \caption{A table to summarise the relation between Eu abundance observation and halo properties.}
    \label{tab:halo distinguish}
\end{table}
\subsection{NSMs or CCSNe?}
    We have assumed that NSMs are the only r-process element sources. There are also other suggested sources and mechanisms: magneto-rotational SNe and collapsars \citep{Woosley93_collapsar, Siegel19_collapsar, Nishimura15_MRSNe}, which are rare types of CCSNe.
    NSMs are different from the SNe-origin mechanisms in that 
    NS binaries need time to merge after formation and that they can travel over long distances before merging.
        
    Interestingly, recent studies on Galactic chemical evolution studies emphasize
    the difference in delay time to conclude
    that rare SNe are favored as the r-process element source than the NSMs. \cite{van_de_Voort19_MWRprocess} argue that rare, special CCSNe with r-process yields comparable to the NSMs is favoured as the main r-process source over the NSMs. The key feature is the flat trend observed in \EuoFe - \FeoH\ plot in metal-poor regime.
    Models with NSMs with long delay times do not make stars to have low \FeoH, but rare, special CCSNe,
    can synthesise and mix r-process elements with the ambient gas quickly so that
    stars formed from the enriched gas have low [Fe/H]. This is a notable feature different from UFDs.
    
    Tuc~III is moderately enriched with r-process elements.
    In our simulations, the best-reproducing scenario is the one with explosion with the offset of a virial radius (see Fig.~\ref{fig:Tuc III scatter plot}), and all central explosion scenarios overproduce r-process elements.
    The relatively low Eu abundance in Tuc~III stars can have three different interpretations: (i) explosion happens at outside the galaxy, therefore only a small fraction of Eu is captured in the stars, (ii) mixing gas mass is quite large, therefore \EuoH\ is small, or (iii) the r-process yields is small in the first place. We argue that the first scenario is the most likely. 
    As for the second argument, we need $4\times 10^{7} \Msun$ of hydrogen gas to reproduce $\EuoH \simeq -2.0$. This is too massive for UFD progenitors at $z>6$, because the typical halo mass
    is $10^{7} \sim 10^{8} \Msun$, and the baryon fraction is only about 16 per cent of all matters. Such high mixing mass is unlikely for the UFD progenitors. This mixing mass estimate is degenerated with the Eu yield from an NSM. A smaller Eu yield makes it possible that Tuc~III Eu abundance is explained by large mixing mass.
    In order to constrain this scenario further, we need UFD observations with lower \EuoH\ abundances.
    As for the third argument, i-process in, e.g. magneto-rotational supernovae (MRSNe) \citep{Nishimura17_iprocess}, is suggested as the small-yield event. Such a weak r-process event has a relatively high electron fraction, and produces a large amount of first-peak elements compared to heavier elements such as lanthanides. However, Tuc~III is known to be lanthanide-rich \citep{Ji19_Lafraction}, suggesting the progenitor r-process event had a quite low electron fraction. Since we do not know any such a neutron-rich and low r-process yield event, we argue that the low r-process yield scenario is unlikely, although it is not rejected.
    
    Interestingly, the other UFDs do not contain Eu-enriched stars,
    but some of them contain very small amounts of Ba and Sr, which probably originate from the r-process \citep{Ji19_Ba_Sr_UFD}. As shown in Fig.~\ref{fig:Overall abundance},
    if a NSM occurs outside the virial radius, it is possible to enrich
    the ISM in star-forming regions to a very low level.
    The stars in UFDs except for Ret~II and Tuc~III typically have [Ba/H]$\ \sim -4$ \citep{Ji19_Ba_Sr_UFD}. Assuming that the ``pure'' r-process produces [Ba/Eu]$\ = -0.89$ \citep{2000Burris_pure_rprocess},
    this corresponds to [Eu/H]$\ \sim -3.1$.
    We have examined our models with halo1, and have found that the low abundance is realised
    if the NSM occurs very far from the center, about three times the virial radius of the simulated galaxy.   
    We have shown that the low Eu abundance of Tuc~III is explained by a NSM in the outskirt
    of the galaxy. If future observations discover UFDs with lower (but non-zero) \EuoH,
    NSMs with large initial kick velocities are strongly favored over other prompt enrichment
    processes.
    
\subsection{Other implications}
From the Eu and Fe abundances in a UFD, we can infer the explosion site and the time lag between
the NSM and typical star formation in the galaxy.
For example, if a large scatter is found in the abundances of
r-process elements, it is likely that
there was long star formation over a few hundred million years after the NSM.
Table~\ref{tab:halo distinguish} summarises our findings.

Other properties such as the explosion energy of the NSM do not affect the Eu abundances dramatically.
We naively expect that NSMs with higher explosion energy results in efficient mixing of the ejecta.
However, since the dominant mixing mechanism is the turbulent motions in the galaxy,
the NSM explosion energy causes a relatively minor effect.
The Eu abundance scatter is determined by the time lag between NSM and star formation quenching,
we do not expect that the merger delay time is important for the [Eu/Fe] - [Fe/H] plot.
\cite{Safarzadeh17_UFD} examine the cases with three different explosion energies, three different delay times, and two different merger sites. They conclude that the merger site is most important in determining the \EuoFe\ - \FeoH\ plot.
Our result is consistent with this notion.
We have also seen that we cannot distinguish explosions at the galactic centre and at 0.1kpc distant from the centre
(Figure \ref{fig:Overall abundance}).

It has been suggested that the r-process enhanced halo stars 
in Milky Way originate from disrupted UFD-like galaxies.
\cite{2018Naiman} reproduce the overall [Eu/Fe]-[Fe/H] trend in the stars of Milky Way by post-processing sub-grid mixing of r-process elements. They find that the [Eu/Fe] distribution of in-situ and ex-situ stars are similar to each other, 
suggesting that the UFD-disruption scenario may not be favoured. Note, however, that they also report paucity of highly r-process enhanced ([Eu/Fe] $\gtrsim 1.5$) stars.
\cite{Safarzadeh19_MWRprocess} study the origin of r-process elements in Milky Way by comparing in detail the distribution and abundances of r-process enhanced metal-poor stars.
Their simulations resolve small progenitor galaxies ($\sim 10^{8} \mathrm{M}_{\odot}$), and the effective mixing timescale in their model is 
consistent with what we find in the present paper.

It appears necessary to resolve small ($\sim 10^{8} \mathrm{M}_{\odot}$) progenitor galaxies to form highly r-process enhanced stars. Such stars are also reproduced in our simulated UFDs, which may constitute, after tidal disruption, a part of the Milky Way halo.

Finally, we discuss possible implications of our results for r-process enrichment 
in globular clusters (GCs). It is known that at least half of GCs have
stars that contain r-process elements \citep{Roederer11_GCrprocess}.
The origin is not identified, and theoretical work is ongoing.  \cite{2019Zevin_GC} models the r-process enrichment of the GCs. They conclude that NSM
can enrich the GCs if the GC contain star-forming gas $30 \sim 50$ Myr after the initial star formation. There is also a curious fact 
that the stars in GCs have very small abundance scatters \citep{Bekki17_GCprolongedSF}. 
Although the UFDs we have considered are 
more massive and physically extended systems than GCs,
our model may offer an interesting scenario that sufficient mixing
of r-process elements through prolonged star formation
causes nearly uniform r-process abundance.
If a fraction of GCs had small galaxies as their progenitors, and if stochastic 
events like NSMs dispersed r-process elements within their progenitors,
it may be possible to reproduce the observed abundance patterns in GCs.

\section{Conclusions}
We have performed cosmological simulations 
to study the physical processes that shape the stellar Eu abundance distribution of UFDs.
We have focused on the explosion sites in the galaxy and star formation histories.

Our results suggest that 
\begin{enumerate}
\item For both Ret~II and Tuc~III, the r-process element abundances can be explained by the NSM model.
  
    \item If a NSM occurs inside the virial radius of a UFD progenitor before reionisation,
      it can enrich the galaxy with r-process elements up to \EuoH $\ \sim -0.5$, depending
      on the explosion site and the gas mass in the galaxy.
      This conclusion also holds for other, similar enrichment processes such as a particular type of
      CCSNe with the explosion energy of $10^{51}$ erg and $\sim$ 0.01 \Msun \ of r-process materials.

    \item The Eu abundance pattern of Ret~II is reproduced if a NSM occurs inside virial radius
      and star formation in the galaxy lasts over a few hundred million years afterwards.
      The abundance pattern of Tuc~III can be explained by the NSM at around virial radius,
      and if star formation lasts for a similarly long time.
    \item With \EuoH\ and \EuoFe, we can estimate the time gap between explosion events
      and typical star-formation in the galaxy.
\end{enumerate}

UFDs are thought to have born early in the history of the Universe. The origin of r-process
elements and how the elements were dispersed in such early galaxies
have been largely unknown, but NSMs are emerging as a promising candidate
especially after the multi-wavelength observations of 
GW170817 and its electromagnetic counterpart.
In the future, a concerted study on stellar populations in galaxies,
physical properties of NS binaries, and the occurrence rate of gravitational waves
by NSMs will reveal how r-process elements were produced in UFDs and also in our Galaxy.

\section*{Acknowledgements}

We thank Volker Springel for kindly providing the simulation code \textsc{arepo}.
We thank Yutaka Hirai, Kenta Hotokezaka for fruitful discussions. This study was supported
by World Premier International Research Center Initiative (WPI), MEXT, Japan and by SPPEXA through JST CREST JPMHCR1414. The numerical computations presented in this paper were carried out on Cray XC50 at Center for Computational Astrophysics, National Astronomical Observatory of Japan.
SI receives the funding from KAKENHI Grant-in-Aid for Young Scientists (B), No. 17K17677.




\bibliographystyle{mnras}
\bibliography{rprocess}







\bsp	
\label{lastpage}
\end{document}